\newcommand{\la}{\langle}
\newcommand{\ra}{\rangle}

\newcommand{\beq}{\begin{eqnarray}}
\newcommand{\eeq}{\end{eqnarray}}

\newcommand{\mn}{{\mu\nu}}

\renewcommand{\theequation}{\thesection.\arabic{equation}}

\newcommand{\btem}{\bibitem}
\newcommand{\TH}{T.\ Hatsuda}
\newcommand{\TK}{T.\ Kunihiro}
\newcommand{\PL}{Phys.\ Lett.\ {\bf B}}
\newcommand{\PTP}{Prog.\ Theor.\ Phys.}
\newcommand{\PR}{Phys.\ Rev.}
\newcommand{\PRL}{Phys.\ Rev. \ Lett.}
\newcommand{\NPB}{Nucl.\ Phys.\ {\bf B}}
\newcommand{\NP}{Nucl.\ Phys.\ }

\newcommand{\HK}{T. Hatsuda and T. Kunihiro}
\newcommand{\KH}{T. Kunihiro and T. Hatsuda}

\documentstyle[12pt]{article}

\begin{document}

\hfill UTHEP-270,\ \  RYUTHP 94-1

\hfill January 1994

\bigskip

\begin{center}

{\Large {\bf QCD Phenomenology }}\\
{\Large {\bf based on a Chiral Effective Lagrangian}}\\

\vspace{1.0cm}

  Tetsuo Hatsuda  and Teiji Kunihiro$^{\ast}$\\

\bigskip
  Institute  of Physics,  University of Tsukuba, Tsukuba, Ibaraki 305,
  Japan\\
  $^{\ast}$Faculty of Science and Technology, Ryukoku University, Seta,
 Otsu-city \ \ \ \ \\
520-21, Japan
\end{center}

\vspace{1.2cm}

\abstract
    We review the Nambu-Jona-Lasinio (NJL) approach to the dynamical breaking
 of chiral
 symmetry in Quantum Chromodynamics (QCD).
    After a general overview of
 the non-perturbative aspects of QCD, we introduce  the NJL model
 as a low-energy effective theory of QCD.
  The collective nature of hadrons and the constituent quark model
  are treated in a
 unified way. Various aspects of QCD related to the
 dynamical and explicit breaking of chiral symmetry and the axial anomaly
 can be well described.
 The subjects treated in Part I include the vacuum structure of QCD, mass
 spectra and coupling constants of hadrons, flavor mixing in mesons,
 the violation of the OZI rule in baryons,
 and the validity of the chiral perturbation in QCD.
 A subtle interplay between the axial anomaly and the current-quark masses
 is shown to play important roles, and a realistic
 evaluation of the strangeness and heavy quark contents of hadrons is given.
   Also the problem of
 elusive scalar mesons is studied in detail.
 For a  pedagogical reason, we first present
  an  account of  basic ingredients and  detailed technical aspects
   of the NJL model using simple versions of it.

In Part II, the NJL model is applied to the system at finite temperature ($T$)
 and density ($\rho$) relevant to the early universe, interior of the
 neutron stars and
 the ultra-relativistic heavy ion collisions.  After a brief introduction
 of the field theory
 at finite temperature,  phenomena associated with the restoration of chiral
 symmetry
 in the medium are examined.  The subjects treated here include the
 quark condensates
 in the medium, meson properties at finite $T$ ($\rho$) and their experimental
 implications.
  A special attention is paid to
 fluctuation phenomena  near the critical
 temperature, i.e.,  possible existence of  soft modes  in the scalar
 channel and a jump of the quark-number susceptibility in the vector channel.

\vspace{1cm}

\hfill to be published in Physics Reports

\hfill Full text and figures are available as hardcopies on request.

\newpage

\begin{flushleft}
Contents:
\end{flushleft}

\vspace{.5cm}

\begin{center}
{\bf Part I}
\end{center}

\begin{enumerate}
\item Overview
 \begin{description}
  \item[1.1] Introduction
  \item[1.2] Basic aspects of QCD
  \item[1.3] Vacuum structure of QCD
  \item[1.4] Dynamical quark mass and chiral symmetry breaking
  \item[1.5] Effective theories of QCD
  \item[1.6] Nambu-Jona-Lasinio model
  \item[1.7] Relation to the other approaches
  \item[1.8] Related review articles
 \end{description}
\item Basic Ingredients of the NJL Model
  \begin{description}
   \item[2.1] The NJL model with one flavor
     \begin{description}
      \item[2.1.1] Lagrangian
      \item[2.1.2] Mean-field approximation
      \item[2.1.3] Determination of the vacuum --- the gap equation and the
                   effective potential
      \item[2.1.4] Decomposition of the Lagrangian into the mean-field
                   and the residual interaction terms
      \item[2.1.5] Dispersion equation for meson spectra; relativistic
                   self-consistent mean-field theory
      \item[2.1.6] Approach based on effective action
      \item[2.1.7] Cutoff scheme
      \item[2.1.8] Meson masses in the chiral limit
      \item[2.1.9] Soft modes and the tachyon pole in the Wigner phase
      \item[2.1.10] Quasi-spin formalism
     \end{description}
   \item[2.2] Two-flavor case
     \begin{description}
      \item[2.2.1] The gap equation and the dispersion equations
      \item[2.2.2] Some physical quantities and current-algebra relations
      \item[2.2.3] Determination of the cutoff and the coupling constant
     \end{description}
 \end{description}
  \begin{flushleft}
   Appendix 2.A\\
   Appendix 2.B\\
   Appendix 2.C\\
   Appendix 2.D
  \end{flushleft}
\item The NJL model with 3-flavors
  \begin{description}
   \item[3.1] NJL Lagrangian with the anomaly term
   \item[3.2] Mean-field approximation
   \item[3.3] Determination of the parameters
   \item[3.4] Physical quantities and the chiral quark model
   \item[3.5] $SU_V(3)$ breaking and the validity of the chiral perturbation
   \item[3.6] Flavor mixing in $\eta$ and $\eta'$ mesons
     \begin{description}
      \item[3.6.1] The interaction Lagrangian
      \item[3.6.2] The masses and the mixing angle
      \item[3.6.3] Coupling of quarks and nucleons to $\eta$ and $\eta'$
      \end{description}
   \item[3.7] Scalar mesons
     \begin{description}
      \item[3.7.1] Masses and the mixing angle
      \item[3.7.2] Where are the scalars?
      \item[3.7.3] Origin of the different mixing in the $\sigma$ mesons and
                   the $\eta$ mesons
     \end{description}
   \item[3.8] Summary and remarks
  \end{description}
  \begin{flushleft}
 Appendix 3.A
  \end{flushleft}
\item OZI violation in Mesons and Baryons --- A Unified Approach ---
  \begin{description}
   \item[4.1] $\Sigma_{\pi N}$ and the $\bar ss$ content of the nucleon
     \begin{description}
      \item[4.1.1] From current quarks to constituent quarks
      \item[4.1.2] Anti-screening of the scalar charge
     \end{description}
   \item[4.2] Light quark content of octet/decuplet baryons
   \item[4.3] Non-linear $m_s$-dependence
   \item[4.4] Intrinsic heavy-quarks in light hadrons
     \begin{description}
      \item[4.4.1] Trace anomaly and intrinsic heavy quarks
      \item[4.4.2] HIggs-baryon coupling constants
     \end{description}
   \item[4.5] Quark and gluon contents of the pion and the kaon
   \item[4.6] Summary
  \end{description}

\begin{center}
{\bf Part II}
\end{center}

\item Introduction to Part II and Primer to Field Theory at Finite Temperature
  \begin{description}
   \item[5.1] Introduction
   \item[5.2] Primer to Field Theory at Finite Temperature
     \begin{description}
      \item[5.2.1] Green's functions
      \item[5.2.2] Correlation functions and susceptibilities
      \item[5.2.3] Variational method for order-disorder transition
     \end{description}
   \end{description}
\item Chiral Restoration and Character Change of Hadrons
  \begin{description}
   \item[6.1] Effective Lagrangian
   \item[6.2] Thermodynamical quantities
   \item[6.3] Mesonic excited states
     \begin{description}
      \item[6.3.1] Dispersion equations
      \item[6.3.2] Coupling constants; chiral algebraic relations at $T\not=0$
      \item[6.3.3] Width of $\sigma$
      \item[6.3.4] Numerical results for meson properties at finite temperature
     \end{description}
   \item[6.4] Discussion
   \item[6.5] Summary
 \end{description}
  \begin{flushleft}
  Appendix 6.A
  \end{flushleft}
\item Soft Mode Associated with Chiral Transition
  \begin{description}
   \item[7.1] Notion of soft modes
   \item[7.2] Elementary excitations in hot QCD
   \item[7.3] Soft mode in the NJL model
   \item[7.4] Hadronic correlations on the lattice
   \item[7.5] Concluding remarks
  \end{description}
\item Quark-number Susceptibility and Vector-Mesonic Modes at High Temperature
  \begin{description}
   \item[8.1] Quark-number susceptibility; the problem
   \item[8.2] Free quark gas
   \item[8.3] Fluctuation-dissipation theorem
   \item[8.4] Model calculation
   \item[8.5] Summary and concluding remarks
  \end{description}
\item Concluding Remarks and Outlook
\end{enumerate}
\begin{flushleft}
References
\end{flushleft}

\newpage
\begin{center}
{\large{\bf Part I}}\\
\end{center}

\bigskip

\section{Overview}
\subsection{Introduction}
\setcounter{equation}{0}
\renewcommand{\theequation}{\arabic{section}.\arabic{equation}}

The dynamics of mesons and baryons are now believed to be described by the
theory of quarks and gluons, i.e.,
 the quantum chromodynamics (QCD) \cite{QCD}.
  The final
goal of the study of QCD is to classify all the hadronic phenomena in terms of
the quark-gluon degrees of freedom. For very high energy
 processes such as the deep
inelastic lepton-hadron scattering (DIS), this program has
 been achieving remarkable success,
 which is due to the asymptotic-free nature of QCD;
 the perturbation theory works in the high energy regime \cite{MUTA}.
 On the other hand, at low energies comparable to the low lying hadron
masses ($\sim 1$ GeV),
 QCD shows  non-perturbative behaviors  such as the confinement of quarks
and gluons and the dynamical breaking of chiral symmetry,  which make the
analytic study of QCD very difficult.

Nevertheless, with  the accumulated  experimental information,
 we have now a good number of
 empirical facts at low energies
 which  are waiting to be explained in a unified manner,
  and thereby give us  hints for making a  unified picture of the
non-perturbative  nature of QCD.   Some examples are:

\noindent
(i) {\em Non-perturbative vacuum structure} \cite{SHUR82}:
 Unlike the QED case, the QCD vacuum
 has non-perturbative condensations of quarks and gluons.
 This fact  is extracted from the studies
 of the current algebra, QCD sum rules and the
lattice QCD.  However, we do not yet know the essential physical
 mechanism of such condensations.

\noindent
(ii) {\em Existence of the Nambu-Goldstone (NG) bosons and
 dynamical breaking of chiral symmetry (DBCS)} \cite{PAGELS}:
The existence of the light pseudo-scalar mesons (pion, kaon and $\eta $)
  suggests that they are the
 NG-bosons associated with the dynamical breaking of
 $SU_L(3) \otimes SU_R(3)$ chiral symmetry down to $SU_V(3)$.
 The pair condensation $\la \bar{q}q \ra$
 in the vacuum deduced from the current algebra relations also supports
this picture.

\noindent
(iii) {\em $U_A(1)$ anomaly} \cite{CHRIS,THOOFT86}:
 The heavy pseudo-scalar particle
 $\eta'$ and also the anomalous decay $\eta \rightarrow 3 \pi$
 suggest  that the $U_A(1)$ quark-current  is not conserved
 even partially.
 This explicit breaking of $U_A(1)$ symmetry in the quantum level
 has a close connection to a topological nature of
the QCD ground state.

\noindent
(iv) {\em Explicit $SU_V (3)$ breaking}:
 The hadron mass differences in the same
$SU_V(3)$ multiplet look as if  being  governed by the
linear breaking of the $SU_V(3)$ symmetry which is summarized as the
Gell-Mann-Okubo (GMO) mass formula \cite{GMO}.
 Although such $SU_V(3)$ breaking is dictated with the current quark masses
 in the QCD Lagrangian ($m_{u,d} \sim O(10$) MeV for $u$ and $d$ quarks
and $m_s \sim O(200)$ MeV for the strange quark),
 the  relation of the quark mass differences
 and those of the  hadron masses  is not simple.

\noindent
(v) {\em Success of the constituent quark model}:
 Low lying mesons and baryons except for NG-bosons are well
described as  systems composed of massive $u$, $d$ and $s$ quarks with
 $M_{u,d} \sim 300 $ MeV and
 $M_s \sim 500$MeV.  These masses
 are determined so as to reproduce the magnetic
moments of the  octet baryons.
 In this constituent quark picture, the baryon mass splittings
 are governed by the $SU_V(3)$ breaking due to the quark mass
difference ($M_s \neq M_u $), the Fermi-Breit type spin-spin interaction
 between the constituent quarks and the orbital excitations of the
 quarks inside the baryons \cite{CQM}.
 The GMO mass formula and its
generalization are obtained successfully  in this picture:
 One of the remarkable example is the $\Sigma - \Lambda$ mass splitting which
 is due to  $M_s \neq M_u$ and the spin-spin
interaction.

\noindent
(vi) {\em OZI rule in hadrons}:
  The flavor changing processes are suppressed in most
of the mesons (such as the $J/ \Psi $ decay into light hadrons,
 small $\omega - \phi$ mixing etc)
 which is summarized as the Okubo-Zweig-Iizuka (OZI) rule \cite{OKUBO}.
  There are, however, several exceptions:
 A  large flavor mixing in the iso-singlet NG bosons
 ($\eta $ and $\eta'$) is known from the 2-photon decays \cite{eta}, for
 instance.
 The non-negligible contents of strangeness and heavy quarks in the  nulceon
 are suggested from
 the analysis of the pion-nucleon sigma term ($\Sigma_{\pi N}$) \cite{JK},
 the spin-dependent structure function in the deep inelastic
scattering \cite{EMC} and the heavy quark production in high energy
proton-nucleus reactions \cite{fermi}.

 Each empirical fact has so far been studied  separately;
 a  physical picture to unify them   has not  been presented yet.
 The lattice simulation of QCD, which is supposed to be the
most fundamental approach for such
non-perturbative problems, is unfortunately not mature enough to give us a
clear physical idea.

One should also note
that it is an intersting subject to study how these aspects of QCD
change at finite temperature ($T$) and finite baryon density ($\rho$).
 Such a study is quite relevant to the physics of the ultra-relativistic
 heavy ion collisions,  the early universe
 and   the neutron stars \cite{QM92}.

In this report,
 we try to give a qualitative description of the important aspects of QCD such
 as (i)-(vi) and the hot/dense QCD
 in a unified way with the use of an
  effective theory of QCD at low energies.
 Our Lagrangian takes a form of the Nambu-Jona-Lasinio (NJL) type \cite{NJL}
 and  has three independent ingredients, i.e.,
 DBCS, $U_A(1)$ anomaly and the explicit symmetry breaking
 due to the current quark masses. Various empirical aspects of QCD are
 then described as a mutual
interplay among the three ingredients.
   Furthermore,
the simultaneous description of the ground state and the collective
excitations in the NJL model allows us to study the change of hadron
properties in hot/dense medium in a self-consistent way.

 We will  introduce a NJL-type 4-fermion interaction between
 the massless quarks with $U_L(3) \otimes U_R(3)$ chiral invariance to
  describe DBCS.  The interaction has a strong attractive
 force between quark and
anti-quark in the $J^p$=0$^+$ channel.
  This induces the instability of the
Fock vacuum of massless quarks to realize  the non-perturbative
 ground state with
 $\bar{q}q$ condensation, which is quite similar to the BCS
mechanism of the superconductivity \cite{BCS}.
  Due to the pair condensation, the original
symmetry of the theory is broken down to $U_V(3)$ and the dynamical quark mass
 $M$ is generated.  The mass  $M$ is nothing but
a quantity which should be identified with the
constituent quark mass in the constituent quark model. Actually, the NJL model
 predicts $M$ around 300 (500) MeV for $u, d$ quarks($s$ quark).
  Thus the notion of
chiral symmetry and the success of the non-relativistic quark model is
reconciled.
 After DBCS, there still remains  residual interactions between
  the constituent quarks, which  give rise to
 collective excitations, i.e., mesons in the new vacuum.
  In the pseudo-scalar
 channel, the strong  residual force makes  the massless NG
bosons, while  in the scalar channel,   scalar bosons
(the chiral partners of NG bosons) with the mass $2M$ are formed.
 The simple relation for the mass-ratio $m_{NG}$ : $M$ : $m_{scalar}$=0:1:2
 is called the Nambu-relation \cite{NJL,NAMBU}.
  Up to this point, there is no flavor mixing among different flavors
 if only the  4-fermion interactions are adopted
 (decoupling of flavors).

 As a next ingredient, we will introduce
 the  current quark masses, which immediately
 causes  mass splittings among the mesons (and baryons)
 with different flavors.   In particular,
 $u \bar{u}$, $d \bar{d}$ and $s \bar{s}$ are preferred as the
 mass eigenstates due to a combined effect of
 the quark-masses and the decoupling of flavors in the interaction.
 Thus the realistic composition
 $\pi^0 \sim u \bar{u} - d \bar{d}$,
$\eta \sim u \bar{u} + d \bar{d} - 2 s \bar{s}$ and
$\eta' \sim u \bar{u} + d \bar{d} + s \bar{s}$ are not yet realized at this
level, while other particles are sitting almost in the right places.

To describe baryons, we will utilize a constituent quark model with the
 quark mass identified with  $M$ introduced above.
  The resulting masses are in  excellent agreement with the
 empirical values, and hence satisfy the celebrated GMO mass formula and
 the equal-mass spacing rule.
 Nevertheless,  the strange quark mass $m_s \sim O(200)$ MeV
 turns out to be too large
  as an expansion parameter in QCD  in contrast with $m_{u,d}$.
 In other words,  the success of the GMO mass-formula
  does not imply the validity of the chiral perturbation
 (expansion by $m_s$). This observation  resolves the
 wide-spread confusion about the magnitudes of  the OZI violation in baryons.
 (See Chapter 3).

 Finally,  a term simulating the $U_A(1)$ anomaly, which we  shall call
Kobayashi-Maskawa-'t Hooft (KMT) term \cite{KOMAS,THOOF}, is introduced.
 The KMT term is a 6-fermion interaction written in a determinantal form
 and gives rise to a mixing between  different flavors.
  On the contrary,  the effect of
large $m_s$  tends to suppress the mixing of ($u$, $d$)
 quarks with $s$ quark, as noted above.
   Thus various observables related to the flavor mixing
 are the results of an interplay between the two effects.
 For example, in the
pseudo-scalar channel, the KMT term dominates over
 the effect of $m_s$. This
 causes  a large flavor mixing among iso-singlet
 NG bosons towards   the right composition
  ($\pi^0 \sim u \bar{u} - d \bar{d}$,
 $\eta \sim u\bar{u} + d \bar{d} - 2 s \bar{s}$ and
 $\eta' \sim u\bar{u} + d \bar{d} + s \bar{s}$).  Simultaneously, the
 $\eta'$ mass is pushed up by the KMT term,
 hence  the resolution of  the $U_A(1)$ problem.
 In the scalar channel, the situation is found
 opposite and the $s$ and ($u$,$d$) do
 not mix so much.  Nevertheless the small mixing still remains
 and shows up  in  the $\bar{s}s$ content of the nucleon.
  By fixing the strength of the KMT term in the $\eta-\eta'$ sector,
 one can thus calculate  non-valence components of baryons, at least
 those in the scalar channel.
  Furthermore, by combining the results with the heavy quark mass expansion in
QCD, one can evaluate  heavy-quark contents of mesons and baryons which
are relevant to the heavy-quark production experiments.

 It is in order here to mention  the significance of the scalar meson $\sigma$
 (chiral partner of the pion) in QCD.
 A model-independent consequence of DBCS is the
 existence of the pion and its chiral partner ($\sigma$):
 The former is the phase
fluctuation of the order parameter $\bar{q}q$
 while the latter is the amplitude fluctuation of $\bar{q}q$.
 The importance of $\sigma$ has been overlooked so far since the
 existence of such a resonance is obscure  in the
 $\pi$-$\pi$ scattering phase shift in the $I=J=0$ channel below 1
GeV \cite{eta}.
 On the other hand,
 light $\sigma$ below 1 GeV is suggested theoretically \cite{SCAD}:
  The quark version of the NJL model gives
 $m_{\sigma} \simeq 2 M \simeq 700$ MeV \cite{HK1f} as
  a result of the Nambu-relation,
 and the mended symmetry by Weinberg predicts
 $m_{\sigma} \simeq m_{\rho} \simeq 770$ MeV \cite{WEIN}.
  Even the lattice data (taking into account only the connected diagram
 though) shows rather low-mass scalar meson \cite{SCALAR}.

 The apparent contradiction between the theory and experiment can be  resolved,
 if the $\sigma \rightarrow 2 \pi$ decay width is so large
  and  comparable to $m_{\sigma}$.
 This picture is again qualitatively
supported by the NJL model (and also in the linear $\sigma$ model) where
 the  decay width is  calculable without introducing free parameters.

Now once the temperature of the system is raised, the $\bar{q}q$ pair
condensation tends to be broken by the thermal fluctuation.
 In the NJL model in the chiral limit ($m=0$),
 the condensation disappears above $T_c \sim (150-200)$ MeV,
 which indicates that
 the restoration of chiral symmetry occurs above $T_c$.
  Associated with this phase transition, collective excitations,
 in particular, the phase fluctuation (pion) and the amplitude fluctuation
($\sigma$) change their properties.  As $T \rightarrow T_c$,
 $\sigma$ tends to
degenerate with $\pi$; thus $m_{\sigma}$ decreases and so does its width
 due to the depletion of the phase space.
 This means that the $\sigma$-meson, which is a broad resonance in the vacuum,
becomes a good  elementary excitation with a small width
 near $T_c$ \cite{HK2a,HK2d}.  Another interesting
observation is that, even above $T_c$, there may exist
  a large fluctuation of the
order parameter with
 the  quantum numbers of $\pi$ and $\sigma$ \cite{HK2a,HK2d}.
 The  modes are analogous  to those corresponding to
 the fluctuation of the order parameter in the superconductor above $T_c$,
 which is responsible for  an anomalous increase
 of the current conductivity
 (Maki-Aslamazov-Larkin mechanism) \cite{MAL}.  Because of their
 small mass and width, they should be   as good
 elementary excitations  as  quarks and gluons above $T_c$.
 Recent lattice QCD simulations of the hadronic modes above $T_c$ seem
 to support this picture \cite{latticeT}.

The restoration of chiral symmetry under the strong background field (either
electromagnetic field or color field) is also an interesting problem.
Since an extensive review on this subject has been published
  \cite{KLEV}, we will not pursue this problem
in this report.

Our effective Lagrangian (the NJL  model)  embodies essential ingredients
of QCD at low energies with a few parameters and is workable but admittedly
 not equivalent with the whole QCD dynamics.
 An advantage of such
 approach lies in the fact that one can easily test  physical ideas by
 rather simple calculations and also
 get possible phenomenological consequences of the basic ingredients of QCD.
  Our intention is  not to give a precise quantitative description of the data
within a few \% level.  Instead, we are interested  in  developing
  physical ideas toward getting a unified picture
 and in calculating observables which would be  hard to do
 in other approaches.  In this sense, our
effective theory written with  quark fields is an intermediate theory between
 the QCD Lagrangian and the low energy chiral Lagrangians
 written in terms of only meson fields with
 infinite number of coupling constants.
Once one  gets qualitative ideas from our theory, one can check them by
 new laboratory experiments or numerical experiments (lattice QCD).

In the following sections of this chapter, we will recapitulate the basic
aspect of QCD and also the rules to construct effective Lagrangians at
 low energies. In Chapter 2, we will give  rather detailed
  description of the treatment of  the Lagrangian in simplest cases.
In Chapter 3, the model is applied  to the world with three flavors (up,
down and strangeness), in which the properties of the mesons are examined with
 an  attention to the interplay of the three ingredients above.
 In Chapter 4, baryons are examined in our model, and the relation
 to the constituent quark model are made clear.
 These are Part I of this report.  Part II
 deals with the systems at finite temperature.
 In Chapter 5, we give a general
  introduction of Part II and a brief account of
 field theory at finite temperature.
In Chapter 6, we discuss changes of hadron properties associated with the
chiral restoration.
 In Chapter 7, we explore possible existence of color-singlet
 collective modes above $T_c$ as a precursor of DBCS.
  Chapter 8 is on another aspect of high $T$ phase, i.e., the
 hadronic  correlation
 in the vector channel and the quark-number susceptibility.
 Chapter 9 is devoted to a brief summary and perspectives.

\subsection{Basic aspects of QCD}
\par
   The classical QCD Lagrangian having
 local color $SU(3)$ symmetry is written as
\beq
{\cal L}_{QCD}^{cl} =  \bar{q} (i \gamma_{\mu} D^{\mu} - m_q ) q
 - {1 \over 4} F_{\mn}^a F^{\mn}_a \ \ ,
\eeq
where $q$ denotes the quark field with three colors and $N_f$
 flavors
$q$=($u$, $d$, $s$, $\cdot \cdot \cdot $), $m_q$ is a mass matrix for current
quarks  $m_q$=diag.($m_u, m_d, m_s, \cdot \cdot \cdot $),
 $D_{\mu} (\equiv \partial_{\mu} - i g t^a A_{\mu}^a$)
 is a covariant derivative
with colored gauge field $A_{\mu}^a$ ($a$=1 $\sim $ 8),
 $g$ the strong coupling constant
and $t^a$ the $SU(3)$-color Gell-Mann matrix
  ($[t^a,t^b]=i f_{abc}t^c$,
tr($t^a  t^b$)=$\delta^{ab}/2$).  The field strength $F_{\mn}^a$ is defined
as $F_{\mn}^a = \partial_{\mu}A_{\nu}^a -
\partial_{\nu}A_{\mu}^a + gf_{abc} A_{\mu}^b A_{\nu}^c$.

If one tries to carry out the perturbation theory in terms of $g$ in the
quantum level, one has to introduce the
gauge fixing term and an associated ghost
term obtained from the Faddeev-Popov procedure \cite{FP}.
  Furthermore, because of the
regularization/renormalization procedure, a mass scale $\mu$ (renormalization
scale) comes into the game \cite{COLLINS}.

 The $\mu$-dependence of the renormalized strong coupling constant is
governed by the  $\beta$-function
\beq
\mu {d \over d\mu}g(\mu) = \beta(g) \ \ .
\eeq
As far as $g$ is small enough, $\beta$ is calculable in the perturbation
 theory
\beq
\beta(g) = -{\beta_0 \over (4\pi)^2} g^3 - {\beta_1 \over (4\pi)^4}g^5
  + \cdot \cdot \cdot \ \ ,
\eeq
with
\beq
\beta_0 = 11- {2 \over 3}N_f \ ,  \ \ \
\beta_1 = 102 - {38 \over 3}N_f \ ,
\eeq
where $N_f$ denotes the number of active flavors. Thus one arrives at the
effective coupling constant or running coupling
\beq
\alpha_s (\mu) & = & {g^2 (\mu) \over 4\pi} \\ \nonumber
               &  = &
{12\pi \over { (33-2N_f) \ln (\mu^2 / \Lambda_{QCD}^2 ) } } \cdot
 [1 -  { 6(153-19N_f) \over (33-2N_f)^2 }
 { {\ln (\ln (\mu^2/\Lambda_{QCD}^2))} \over \ln(\mu^2/\Lambda_{QCD}^2)}]
 \\ \nonumber
 &  & + \cdot \cdot \cdot \ .
\eeq
Here $\Lambda_{QCD}$ is a scale parameter characterizing the
change of $\alpha_s$ as a function of $\mu$. It  depends  on the
subtraction scheme and the number of active flavors.
 Analyses of the various high energy processes show \cite{ALT}
\beq
 {\Lambda^{(3)}_{\overline{MS}}}
    &  = & (290 \pm 30) {\rm MeV},  \nonumber \\
 {\Lambda^{(4)}_{\overline{MS}}}
    & = & (220 \pm 90) {\rm MeV},  \nonumber \\
 {\Lambda^{(5)}_{\overline{MS}}}
    & = & (140 \pm 60) {\rm MeV},
\eeq
 where the $\overline{MS}$ scheme \cite{BBDM} is used and
  superscripts indicate the number of active flavors.
 Eq.(1.5) tells us that the running coupling decreases logarithmically as $\mu$
increases and the perturbation theory works well for large $\mu$.
\footnote{This does not
necessary mean that the expansion by $g$ is  convergent for
 large $\mu$.
 Instead, the expansion is  at most asymptotic.
  Large order behaviors of the expansion
 is reviewed in  \cite{WEST}.}  This
 is called the {\em asymptotic freedom} \cite{ASYM}.
  Correspondingly, the running mass obeys the relation
\beq
\mu {d \over d\mu}m(\mu) = - \gamma_m (g) m(\mu) \ \ ,
\eeq
with the mass anomalous dimension \cite{TARR}
\beq
\gamma_m =  2({\alpha_s \over \pi}) + ({101 \over 12} - {5 \over 18}N_f)
 ({\alpha_s \over \pi})^2 + \cdot \cdot \cdot \ \ .
\eeq
The light quark masses determined from the hadron mass splittings and the QCD
sum rules read
\beq
m_u(1{\rm GeV}) & = & (5.1 \pm 0.9) {\rm MeV}, \ \
m_d(1{\rm GeV})=(9.0 \pm 1.6) {\rm MeV}, \\ \nonumber
m_s(1{\rm GeV}) & = & (161 \pm 28) {\rm MeV},
\eeq
from ref.\cite{NAR} and
\beq
m_u(1{\rm GeV}) & = & (5.6 \pm 1.1) {\rm MeV}, \ \
m_d(1{\rm GeV})=(9.9 \pm 1.1) {\rm MeV}, \\ \nonumber
m_s(1{\rm GeV}) & = & (199 \pm 33) {\rm MeV},
\eeq
from ref.\cite{DD}.
Within the perturbation theory, $\mu$ should be chosen so that the higher order
terms in the expansion are effectively
suppressed.  Thus $\mu$ should be a typical scale of
the system.  For example, in the deep inelastic lepton-hadron
 scattering (DIS), the
typical scale is $Q^2=-q^2$ with $q^{\mu}$ being the four-momentum transfer
 to hadrons.
   For the systems at very high temperature ($T$) or at
 high baryon-number density,
  $\mu$ will be identified with $T$ or
 the fermi-momentum $p_{_F}$ of the system \cite{CP,KM}.
This is the reason why we can expect the weakly
 interacting system of quarks and gluons in the high
 temperature/density plasma
(quark-gluon plasma) \cite{QM92}.
  The DIS processes have been investigated in  great detail
in the past two decades and the
violation of the Bjorken scaling predicted by
 QCD was confirmed  \cite{MUTA}.
   As for the quark-gluon plasma (QGP), the
future projects on the ultra-relativistic
 heavy ion collisions at BNL (RHIC project)
and at CERN (LHC project) will shed  light on its nature.

Now, if $\mu$  is
 comparable to $\Lambda_{\overline{MS}}$,  the
perturbation theory does not work anymore and the non-perturbative
 phenomena play  dominant roles, as noted in the previous section.
  Extensive lattice QCD studies \cite{LQCDR},
 although they are still confined in a small lattice volume,
confirm that the low-energy dynamics is really dominated by the
non-perturbative configuration of the self-interacting gluon field.
Unfortunately, because of the limited computer
time and power, one does not
yet reach a truly reliable QCD simulation including
 the virtual $q-\bar{q}$
pair of light quarks.  Furthermore,
 it is rather difficult to extract essential
physics only from the direct numerical simulations. (Several attempts have
been recently started
  to single out the dominant gluonic fluctuations on the lattice \cite{Szhou}.)

\subsection{Vacuum structure of QCD}
\par
 Because of the non-perturbative interactions among quarks and gluons,
the ground state of QCD has a non-trivial structure.
  This structure also affects
the properties of the elementary excitations
 on the ground state, i.e., mesons
and baryons.

 The presense of the non-zero vacuum expectation values of certain
operators is one of the evidences
 of such non-perturbative vacuum structure.
For example, the current algebra relation (Gell-Mann-Oakes-Renner
relation) \cite{GOR} tells us
\beq
f_{\pi}^2 m_{\pi^{\pm}}^2 \simeq
 - \hat{m} \la \bar{u}u + \bar{d}d \ra
\eeq
where $\hat{m}$=$(m_u+m_d)/2$ is an averaged mass of
$u$ and $d$ quarks.  By
taking the known pion decay constant $f_{\pi}$=93 MeV
 and $\hat{m}(1$GeV)=$(7 \pm 2)$ MeV, one gets
\beq
\la \bar{u}u \ra \simeq \la \bar{d}d \ra \simeq
[- (225 \pm 25) MeV]^3 \ \ \ {\rm at}\ \ \mu^2= 1 {\rm GeV} \  \ .
\eeq
This means that the QCD ground state has
condensation of quark and anti-quark
pairs, which is, as noted in a previous section,
  quite analogous to the ground state of the BCS type
superconductor where there is a condensation of the electron Cooper pairs
$\la \psi_{\uparrow} \psi_{\downarrow} \ra \neq 0$ \cite{BCS}.

The analyses based on the QCD sum rules \cite{SVZ}
 for the heavy as well as light quark systems
  show that the gluons also condense in the vacuum \cite{SVZ,RECENT}
\beq
\la {\alpha_s \over \pi} F_{\mu \nu}^a F^{\mu \nu}_a \ra
= (350 \pm 30 \ {\rm MeV})^4 \ \ .
\eeq
The non-zero value of the gluon condensate is
also suggested
 by the  numerical simulations on the lattice \cite{DIG,LEE}.

There are of course lots of other scalar operators which have vacuum
expectation values. One of such operators
with physical significance is the four-quark operator
\beq
\alpha_s \la (\bar{u} u)^2 \ra = (1.8 - 3.8) \cdot 10^{-4} \ \ \ {\rm GeV}^6 ,
\eeq
which
 is essential to determine the masses of $\rho$ and $\omega$ mesons
in the QCD sum rules \cite{SVZ,RECENT}.

Besides the fact that $\bar{q}q$
 and $F_{\mn}^a F_a^{\mu \nu}$ are the low
dimensional operators, they have a
 close connection with the symmetry properties
 of the classical QCD Lagrangian ${\cal L}_{QCD}$.
   Let's summarize here the
 several symmetries of ${\cal L}_{QCD}$.

\noindent
(i) {\em Global chiral symmetry}: Under the transformation
\beq
q_{_L} \rightarrow e^{i\lambda^a \alpha^a} q_{_L} , \ \ \
q_{_R} \rightarrow e^{i\lambda^a \beta^a} q_{_R} ,
\eeq
with $\lambda^0= {\bf 1}\sqrt{2/N_f}$,
 $\lambda^{i} = t^i/2$ ($i=1 \sim N_f^2-1$),
${\cal L}_{QCD}$ has $U_L(N_f) \otimes U_R(N_f)$
 symmetry in the chiral limit ($m_q=0$).

\noindent
(ii) {\em Dilatational symmetry}: Under the scale transformation
\beq
q(x) \rightarrow \epsilon^{3/2} q(\epsilon^{-1}x),
 \ \ \ A_{\mu}^a(x) \rightarrow
 \epsilon A_{\mu}^a (\epsilon^{-1}x) \ \ ,
\eeq
as well as $x_{\mu} \rightarrow \epsilon^{-1} x_{\mu}$,
${\cal L}_{QCD}$ is
invariant except for the quark mass term.

In the quantum level, some of these symmetries are broken by a quantum
 effect (anomaly) even in the chiral limit.  The
 corresponding conservation laws are now modified as
\beq
{\partial_{\mu}(\bar{q} \gamma_{\mu} \lambda^a q)} &  = &
i \sum_{i,j}^{N_f} \bar{q}_i (m_i - m_j)  \lambda^a q_j
\ \ \ (a=0 \sim N_f^2-1) \\ \nonumber
{\partial_{\mu}(\bar{q} \gamma_{\mu} \gamma_5 \lambda^a q)} & = &
i \sum_{i,j}^{N_f} \bar{q}_i (m_i + m_j) \gamma_5 \lambda^a q_j
\ \ \ (a=1 \sim N_f^2-1) \\ \nonumber
{\partial_{\mu}(\bar{q} \gamma_{\mu} \gamma_5 q)} &  = &
i \sum_{i}^{N_f} \bar{q}_i 2m_i \gamma_5 q_i + 2N_f {g^2 \over 32\pi^2}
 F_{\mn}^a \tilde{F}^{\mn}_a  \\ \nonumber
{\partial_{\mu}D^{\mu} = \Theta_{\mu}^{\mu}} & = &
(1+ \gamma_m) \sum_i^{N_f} \bar{q}_i m_i q_i + {\beta \over 2 g}
 F_{\mn}^a F^{\mn}_a  \ \ ,
\eeq
Here the index $i,j$ stand for the flavor,
$\tilde{F}_a^{\lambda \rho}$=
${1 \over 2} \epsilon^{\mn \lambda \rho} F_{\mn}^a$ and $\Theta_{\mn}$
 is the energy momentum tensor of QCD.
The third and last equation show the {\em axial
anomaly} \cite{ABJ} and the  {\em trace anomaly} \cite{ELLIS},
 respectively.

Now the existence of $\la \bar{q}q \ra \neq 0$
 implies  that the $SU_L(N_f) \otimes SU_R(N_f)$ symmetry is broken down
 to $SU_V(N_f)$ i.e., the vacuum breaks the global chiral
 symmetry
$Q_5^a \mid 0 \ra \neq \mid 0 \ra$ \cite{PAGELS}.

Here we note an interesting  physical consequence of the trace anomaly:
 Taking the vacuum expectation value of the last equation of eq. (1.17),
 one can see that  QCD vacuum has a smaller energy   density than the
perturbative vacuum \cite{SHUR82},
 \beq
\epsilon_{vac} - \epsilon_{pert.} =  \la \Theta_{\mu}^{\mu} \ra
 \simeq - \frac 7{8} \la {\alpha_s \over \pi}F_{\mu \nu}^a
F^{\mu \nu}_a \ra < 0,
\eeq
 where a small contribution from the quark condensates is neglected for
 simplicity,
 and $N_f$=6. Here the last inequality is due to eq.(1.13).

\subsection{Dynamical quark mass and chiral symmetry breaking}
\par

 The success of the constituent quark model implies  that
 many of  hadron properties can be described  by a simple
 assumption  that  massive quarks
 with $ M_{u,d} (M_s) \sim 300 (500)$ MeV
  interact rather weakly inside hadrons
  \cite{CQM}.
  Such  large masses of the  constituent quarks are
certainly not the same objects as the current-quark masses
 in the QCD Lagrangian.

  In fact, $\la \bar{q} q \ra$$\neq$0, which we saw in
   the previous section,
suggests the existence of the  "dynamical mass"
in the quark propagator.
 The quark condensate in QCD is given by the trace of the full quark
propagator $S_F$:
 \beq
  \la \bar{q} q \ra = -i \lim_{y \rightarrow x+}
{\rm Tr} S_F(x,y) \ \ .
\eeq
 Since $\bar{q}q$ is a gauge invariant quantity, one can take any
gauge to evaluate $S_F(x,y)$ and it will have a general form
in the momentum space as
\beq
S_F(p) = {A(p^2) \over p \cdot \gamma - B(p^2) }\ \ .
\eeq
At least, within the perturbation theory in the chiral limit,
 there is no
mixing between left handed quarks and right handed quarks. Therefore
$B(p^2)$=0 and the quark condensate never
 takes a non-zero value just because of
a simple trace identity Tr$\gamma_{\mu}$=0.  On the other
hand, $B(p^2)$ can be generated in a
 non-perturbative way, as  shown  originally
 by Nambu and Jona-Lasinio
  in somewhat different context \cite{NJL}
 and also by several authors in the
 approximated version of QCD such as the ladder
  QCD approach \cite{LADQCD}.
  In the NJL model,
\beq
B(p^2) = M \ \ ,
\eeq
which is a result of the self-consistent equation
 (the Schwinger-Dyson equation) with the contact 4-fermi
interaction (See Fig.1.1).  In the ladder QCD,
 $B(p^2)$ has  a momentum
dependence  which is obtained by the infinite sum of the
 rainbow diagram
 (See Fig.1.2).
In both cases, the non-perturbative generation
 of the scalar self-energy (or the
dynamical mass) is intimately related to the generation of
 the non-zero quark condensate in the vacuum.

\

\begin{center}
{\fbox {\bf Fig. 1.1, Fig. 1.2}}
\end{center}

\

Here one might wonder the
 validity of the approximate form of $B(p^2)$ such as (1.21)
 which gives rise  to
  a pole of $S_F(p^2)$. (The QCD vacuum is confining, therefore
there should be no such pole in the exact quark propagator.)
  In fact, the
generation of the dynamical mass is not the end of the story:
 There will be a long-range force between massive quarks
to confine them and also there will be
a short range spin-spin interaction between  massive quarks.
 The former will modify the low momentum part of the propagator to kill
 the on-shell propagation of free quarks.
 This is actually the whole idea of the constituent quark model
 supplemented by the notion of the dynamical symmetry breaking;
 \beq
{H_{QCD}} &  = & H_{{\rm massless} \ {\rm quarks}} +
    H_{\rm massless\ gluons} + V_{int} \\ \nonumber
          & \simeq & H_{{\rm massive}\ {\rm quarks}} +
 v_{int} \ \ ,
\eeq
 where $v_{int}$ is the interaction between massive quarks and
  is expected to be   much weaker than
 the interaction between
bare massless quarks, $v_{int} << V_{int}$ \cite{GOHA}.  In short,
   the massive quarks,
 which are generated by  the chiral symmetry breaking,
 make a  good basis to represent the QCD Hamiltonian at low energies.

Another important observation of the above picture is that
 the conserved quantum
numbers (such as charge,  $SU_V(3)$ flavor quantum numbers etc)
 are the same
for current quarks and the constituent quarks.
  This is essentially due to the
non-renormalization  of the partially conserved currents \cite{JM};
\beq
(\bar{q} \gamma_{\mu} \lambda^a q )_{{\rm low}\ \mu^2}
& = & (\bar{q} \gamma_{\mu} \lambda^a q )_{{\rm high}\ \mu^2}
 \  \ (a=0 \sim 8) \\ \nonumber
(\bar{q} \gamma_{\mu} \gamma_5 \lambda^a q )_{{\rm low}\ \mu^2}
& = & (\bar{q} \gamma_{\mu} \gamma_5 \lambda^a q )_{{\rm high}\ \mu^2}
 \  \ (a=1 \sim 8) \ \ .
\eeq
The currents normalized at
low $\mu^2$ (corresponding to the currents for
constituent quarks) are the same as  those normalized at high $\mu^2$
(corresponding to the currents for current quarks).
Because of this property,
almost all the counting rules and
 the classification based on the $SU_V(3)$
symmetry works for massive quarks as well as
massless quarks, which is again
one of the basic advantages of the quark model.
   One of the well-known
exception of this rule is the
flavor-singlet axial charge of the constituent
quark $g_A^0$.  Since this quantity is
 scale dependent due to the axial anomaly,
  $g_A^0$ of a current quark is different from that of the constituent quarks.

 The constituent quark picture (1.22) has some similarity
with the idea of the Landau theory of the Fermi liquid \cite{Landau}.
 In the Landau theory, the basic properties of the Fermi liquid at
 low temperatures can be accounted for in terms of weakly interacting
 quasi-particles; they  have the
 same quantum numbers as  the bare particles but may have different
  physical properties by wearing clothes  of the interaction adiabatically.
   Our constituent quarks  (current quarks) in QCD at low energies
 are analogous to  the quasi-particles (bare particles) in the Landau theory.

\subsection{Effective theories of QCD}
\par



The basic idea of the effective theories
 is found in many areas of physics:
 Suppose that we are treating the system with many degrees of freedom.
   If we
are only interested in the dynamics
 of a few or small degrees of freedom in the
 total system, we will try to
 derive an effective theory only for the relevant
degrees of freedom by eliminating the irrelevant degrees
 of freedom \cite{EFFECTIVE}.
 The way to eliminate irrelevant variables depends on the system.
   In many cases,
it can be done only in an approximate manner and in some cases, one
does not  even
 have any idea how to do it.
   Unfortunately,  low energy QCD
 is in the last category at the present time.
 In this case, one has to
write down the effective Lagrangian based on
 some general constraints such as
 the symmetry properties.
 In effective theories developed by Weinberg \cite{WEIN2} and later by
Gasser-Leutwyler \cite{GL1}, infinite numbers of coupling constants
 are allowed to describe the nature.
 Our approach is different from theirs and may be  categorized as
 an  intermediate step
between QCD and such approaches  as we will see below.

Our basic clues to write down the low energy effective theory of QCD
are the following:

\noindent
(i) {\em Effective quark theory}:
 we would like to formulate an effective theory
 which can be a basis for the constituent quark model.
  Therefore, the theory is
written in terms of  quark variables without explicit gluons.

\noindent
(ii) {\em Chiral symmetry}: the theory should preserve the same
 global chiral symmetry as
 the QCD Lagrangian does, and  should also
develop the dynamical breaking of the  symmetry.

\noindent
(iii) {\em Low energy degrees of freedom}:
we are interested in the low energy
properties of the quark dynamics
where the energy scale is smaller than
 some cutoff scale $\Lambda \simeq 1$ GeV.
    The short distant
dynamics above $\Lambda$ to be dictated by the perturbative QCD
  will be treated  as a small perturbation.

Thus the Lagrangian of the effective theory is generally written as
\beq
{\cal L}_{eff}(x) = \sum_n c_n {\cal O}_n(x)
({1 \over \Lambda})^{{\rm dim}{\cal O}_n-4}
  \ \ \,
\eeq
where ${\cal O}_n$ are the local chiral-invariant operators
 written with  the
 quark fields and $c_n$ are ther c-number dimensionless
  coupling constants.
 The theory is defined at the scale below $\Lambda$,
  thus the internal loop
momenta of the theory should be cutoff at $\Lambda$.
  The effect of the
higher dimensional operators are suppressed with  the
 powers of $p^2/\Lambda^2$
where $p^2$ is the typical momentum of the system one treats.
 Once we truncates the above series
  and determine the finite number of coupling
constants with  suitable physical inputs,
one can use the theory to
calculate some other quantities.  Note here that the
cutoff scale $\Lambda$
will depends on the number of terms one takes  in the above expansion.

\subsection{Nambu-Jona-Lasinio model}
\par
A simplest effective theory satisfying the requirement in the previous
 section is the Nambu-Jona-Lasinio (NJL) model.
  Originally, the model was
introduced to describe the pion  as a bound state of
 the nucleon and the anti-nucleon \cite{NJL}.
  We will replace their nucleon field by the
quark field
 and adopt only the lowest dimensional operators as ${\cal O}_n$.
Then ${\cal L}_{eff}$ for $N_f=3$ reads
\beq
{{\cal L}_{NJL}} &  = & \bar{q}(i \gamma \cdot \partial - m_q)
  q \\ \nonumber
 &  & + \sum_{a=0}^{8}
 {g_s \over 2}
[(\bar{q} \lambda^a q)^2 + (\bar{q}  i \gamma_5 \lambda^a q)^2] \\
 \nonumber
 &  & + \cdot \cdot  \cdot  \ \ \ ,
\eeq
where $\cdot \cdot \cdot $ denotes
 all the possible 4-fermi interactions with
global $U_L(3) \otimes U_R(3)$ symmetry.

Now in the quark level, we already know that $U_A(1)$ symmetry
is broken by the
axial anomaly.  Since we do not have the explicit gluon field,
 we have to take into
account this effect in the tree level of our quark Lagrangian.
  The lowest
dimensional operator which preserves
   $SU_L(3) \otimes SU_R(3)$ but breaks $U_A(1)$ is
 \beq
{\cal L}_{KMT} = g_{_D} \det_{i,j}[ \bar{q}_i (1- \gamma_5) q_j +
 {\rm  h.c.}] \ \ \ .
\eeq
  This is a dimension 9 term
 introduced in  \cite{KOMAS,THOOF}.
  Eq.(1.25) and (1.26) are
 the basic effective Lagrangian \cite{HK3a}-\cite{HK3g}
 which we are going to examine in detail in this
article.

A main drawback of the NJL model is a lack of confinement.
    However, as we have
 discussed before around eq.(1.22),
 the net effects $v_{\rm int}$ of confinement
  as well as the  short range
interactions
  may be treated as a small perturbation.
This idea was first discussed by Goldman-Haymaker \cite{GOHA}.  In our
 model, DBCS is described by the extended NJL model including
  the KMT term, hence our basic picture may be
 represented as
\beq
{\cal L}_{QCD} \simeq
 {\cal L}_{NJL} + {\cal L}_{KMT}
  + \epsilon ({\cal L}_{conf} + {\cal L}_{OGE}).
\eeq

The effective quark theory for  $N_f$=3 with $U_A(1)$
 breaking term has a long history of
research. The $U_A(1)$ breaking term in the quark level eq.(1.26)
 was first written down by Kobayashi and Maskawa on 1970
 \cite{KOMAS}   to describe the mixing property of $\eta -
\eta'$ and also the heavy $\eta'$ mass.
 After the discovery of the topological
configuration in QCD, i.e., instantons, 't Hooft
derived an effective interaction
for quarks under the instanton background \cite{THOOF}
 which turned out to
be the same structure as  the Kobayashi-Maskawa's
except for an interaction  with an explicit color $SU(3)$ matrix.
 Thus we call (1.26) the  KMT-term.
 Schechter and his collaborators applied the  4-fermi interaction +
 KMT-term to describe the nonet mesons within the long wave length
approximation \cite{MS}.\footnote{The $N_f$=2 version of
${\cal L}_{KMT}$ or its non-local version has
been also studied by several authors \cite{SU2DET}.}
  More serious studies based
 on the random-phase approximation or
 the Bethe-Salpeter equation  have been carried out  by
   Kunihiro and Hatsuda \cite{HK3a}-\cite{HK3g},
 by Bernard, Jaffe and Meissner
\cite{BJM} and by Kohyama, Kubodera and Takizawa \cite{Takizawaa}.
  Alkofer and Reinhardt \cite{AR} developed a useful path-integral
bosonization procedure for the Lagrangian with 6-fermi interactions.
(see also the later extensive studies \cite{Takizawab,WEISE1}).

 The 4-fermion NJL model had been also studied as a
model for extended hadrons.  The pioneering works by Eguchi and Sugawara
\cite{ES} and Kleinert \cite{KL74}  initiated the studies  of
  such theories in the mid 70's \cite{KUGO}.  It was later
 shown that the NJL model reproduces and relates the basic hadronic parameters
  remarkably well   by Ebert and Volkov \cite{EBVOL},
 Volkov \cite{VOL}, and
Hatsuda and Kunihiro
 \cite{HK1f}\footnote{In ref.\cite{HK1a}, the NJL model
 was used  as the Lagrangian describing the
 quark dynamics in the {\em interior} of the chiral bag.}.
  A realistic
   application of the model to the phenomena at finite
 $T$ medium were initiated
  by Hatsuda and Kunihiro \cite{HK1f}-\cite{HK2d} and
 later the systems at finite density
 \cite{HK1f,KOCIC,BMZ,AY}
 and at finite external field \cite{KLEV} have been studied.

 The relation of  the NJL model to the non-linear chiral model due to
 Skyrme, Wess and Zumino, and Witten (SWZW) \cite{SWZW}
  has been also clarified by the
 works of Dhar and Wadia \cite{DW} and Ebert and Reinhardt\cite{ER}.
 The higher derivative terms of the chiral Lagrangian
  compiled by Gasser and
Leutwyler \cite{GL1}
 were successfully reproduced by  simple quark-loop integrals
 and the Wess-Zumino term is
derived as  an imaginary part of the
effective action coming also from  the quark loops.
 (See also ref.\cite{WAKA}).

Applications of the NJL model to the baryons are still
 under active studies.  In this report, we will take a simplest
 constituent-quark picture where the residual interactions
 are assumed to be small (see Chapter 3 for details).
 Other approaches in the NJL model
 include the  diquark-quark model (see e.g.
 \cite{WEISE2}), soliton model (see e.g. \cite{GOEKE})
 and Faddeev approach (see e.g. \cite{FAD}).

It should be also remarked that,
 in the strong-coupling lattice QCD, the
effective quark theory similar to the NJL model  is obtained \cite{KS}.
 This strong-coupling effective theory has been also applied
 to the hot and dense medium \cite{KAWA}.
  There are also some attempts to derive the NJL model
 from the continuum QCD; see e.g.  ref.\cite{BALL}.

\subsection{Relation to the other approaches}

In this subsection, we will give  a brief comment on
the relation between the
 NJL approach and  the other approaches for low energy hadrons.

 Because of the dynamical
symmetry breaking caused by the 4-fermi interaction, the following
 things happen simultaneously in the NJL model;
 non-zero vacuum condensate, generation of the dynamical
quark mass, the creation of the massless Nambu-Goldstone boson
 ($\pi$) and the creation of the chiral partner ($\sigma$).
In the NJL model,  only the quarks are
the basic degrees of freedom and $\pi$ and $\sigma$ are
  $q$-$\bar{q}$ bound states.

 In the path-integral bosonization of the NJL model,
 by introducing the auxiliary fields
  $\pi^a \sim \bar{q} i \gamma_5 \lambda^a q$ and
  $\sigma^a \sim \bar{q} \lambda^a q$, the partition function
  can be written as an integral over
 $q$, $\bar{q}$, $\pi$ and $\sigma$.
   If one further freezes the $\sigma^a$ degrees of
freedom,
 one arrives at an effective theory of quarks and pions
  in the non-linear
liarization.  We will call this
  effective theory  as
"chiral quark model" ($\chi $QM)  which has been
  studied by many authors independently
 of the development of the NJL model \cite{CHQM}.

 One can further integrate out the quark field
   explicitly or approximately and arrive at
 an effective theory written only in terms
of the chiral fields and its derivatives \cite{DW,ER,OTHER}.
 The final form is nothing
 but the SWZW type chiral Lagrangian, which
 contains infinite number of derivative
 terms and also summarizes  the anomalies
 in QCD.

There is one advantageous point of the NJL
 derivation of these effective
Lagrangians. One can calculate all the coupling
 constants between quarks and
 mesons appearing in the chiral quark models
  and all the coupling constants
in the chiral Lagrangian in the SWZW Lagrangian;
  they are controlled only by a few parameters
 in the NJL model.\footnote{This is analogous to the Gor'kov's derivation
 \cite{gorkov}
 of the Ginzburg-Landau theory of the superconductivity \cite{gl}.}
  This fact in turn gives a crucial test
  of  the NJL model
since some of the coupling constants in the chiral Lagrangian
  are known  experimentally.
  We will come to this
point later and we just quote here that within
the 10-20\% level, the NJL
theory agrees to all the empirical facts,
thus quite successful even as a
quantitative theory for hadrons.

Here it is worth-while to  mention the complementary
 roles of the
non-perturbative dynamics at low energies and the
perturbative QCD at high energies.
 Let's write down the dispersion relation for the
 two point function of the bilinear composite operators
  such as $\bar{q} \Gamma q$ with $\Gamma$ being some gamma matrix:
 \beq
 {\rm Re}\Pi (q^2) = {{\rm P} \over \pi} \int_0^{\infty}
 {{\rm Im}\Pi (s) \over {q^2 - s}} ds + {\rm (subtractions)} \ \ .
 \eeq
Since the low energy dynamics below $\Lambda$ is
 assumed to be dominated by ${\cal L}_{NJL}$ and the higher
  energy part is dominated by the perturbative QCD (PQCD),
  Im$\Pi$ has a hybrid form
  \beq
  {\rm Im}\Pi(s) = \theta(\Lambda^2 - s)
  {\rm Im}\Pi_{NJL}(s) + \theta(s - \Lambda^2)
  {\rm Im}\Pi_{PQCD} \ \ \ .
  \eeq
  Thus, as is evident from the dispersion relation,
   Re$\Pi(q^2)$ is a superposition of the high
    energy and low energy contribution.
    The cutoff $\Lambda$ has a meaning of the starting point of the
    continuum threshold of the spectral function or approximately the
    position of the second resonances of the mesons.
     It is interesting that this 4-momentum cutoff  $\Lambda$ turns out to be
     around 1.4 GeV which is quite consistent with the
      analysis of the QCD sum rules and also the locations  of
 the   second excited states of the low lying mesons.

The relative importance of NJL and PQCD depends
 on the region of $q^2$ which one
is looking at.  For low $q^2$ below the cutoff,
 Re$\Pi$ is dominated by the NJL part (i.e. hadronic poles)
 and the PQCD part gives
  only a correction of $O(q^2 / \Lambda^2$).  On the other hand,
   for $q^2$$\rightarrow$$- \infty $,
   Re$\Pi$ is dominated by the PQCD with power
    corrections coming from the NJL part of $O(\Lambda^2 /q^2)$.
  In the latter case, the operator product expansion works for
  Re$\Pi$ and the power corrections are written with the
   various vacuum condensates.  In this sense,
   the NJL model can even give an approximate evaluation of
    such non-perturbative condensates.
    The original work of Sakurai on the finite energy sum
    rule for vector mesons is implicitly based on
   this picture \cite{SAKURAI}.

\subsection{Related review articles}

Since there have appeared several review articles on the NJL model recently,
 we will list them up here for the readers.
 One of the earliest review article is \cite{iiyama} by one of the
present authors. This covers the development before 1989, some parts
 of which have overlap with this article.  Reviews with some emphasis on the
 di-quark picture of the nucleon and the nuclear applications include
 \cite{WEISE2} and \cite{WEISE3}.  \cite{VOL2} pays attention to  the
 meson spectra, and the electro-magnetic and weak interactions of hadrons
 in the NJL model.
  \cite{KLEV} puts emphasis on  the NJL model with  strong external fields.
 The extensive applications of the model to baryons can be seen
 in \cite{GOEKE}.
  These reviews including this article are complementary with each other.

 Since the number of papers published in this field is enormous
 and still developing, it is beyond our ability to search and
 list up all the publications.  We apologize to our
 colleagues if their important contributions are omitted in this
 review.

\end{document}